\documentclass[aps,prb,twocolumn,a4paper,floatfix,showpacs,superscriptaddress]{revtex4}

\usepackage{graphicx}
\usepackage{epsfig}
\usepackage{hyperref}
\usepackage{color}
\usepackage[normalem]{ulem}
\usepackage{float}
\usepackage{multirow}
\usepackage{lineno}
\usepackage{upgreek}  
\usepackage[fleqn]{amsmath}
\usepackage{amssymb}
\begin{document}

	\title{Phase-field simulations of the effect of temperature and interface for zirconium $\delta\mbox{-}$hydrides}
	
\author{Zi-Hang Chen\footnotemark[2]}
\affiliation{School of Materials Science and Engineering,Beijing Institute of Technology, Beijing 100081, China}
\affiliation{Advanced Research Institute of Multidisciplinary Science, Beijing Institute of Technology, Beijing 100081, China}
\affiliation{Laboratory of Computational Physics, Institute of Applied Physics and Computational Mathematics, Beijing 100088, China}

\author{Jie Sheng\footnotemark[2]}
\affiliation{Laboratory of Computational Physics, Institute of Applied Physics and Computational Mathematics, Beijing 100088, China}

\author{Yu Liu}
\email{liu\_yu@iapcm.ac.cn}
\affiliation{Laboratory of Computational Physics, Institute of Applied Physics and Computational Mathematics, Beijing 100088, China}

\author{Xiao-Ming Shi}
\affiliation{Department of Physics, University of Science and Technology Beijing, Beijing 100083, China}

\author{Houbing Huang}
\affiliation{School of Materials Science and Engineering,Beijing Institute of Technology, Beijing 100081, China}
\affiliation{Advanced Research Institute of Multidisciplinary Science, Beijing Institute of Technology, Beijing 100081, China}

\author{Ke Xu}
\affiliation{School of Materials Science and Engineering,Beijing Institute of Technology, Beijing 100081, China}
\affiliation{Advanced Research Institute of Multidisciplinary Science, Beijing Institute of Technology, Beijing 100081, China}
\affiliation{Laboratory of Computational Physics, Institute of Applied Physics and Computational Mathematics, Beijing 100088, China}

\author{Yue-Chao Wang}
\affiliation{Laboratory of Computational Physics, Institute of Applied Physics and Computational Mathematics, Beijing 100088, China}

\author{Shuai Wu}
\affiliation{School of Materials Science and Engineering,Beijing Institute of Technology, Beijing 100081, China}
\affiliation{Advanced Research Institute of Multidisciplinary Science, Beijing Institute of Technology, Beijing 100081, China}
\affiliation{Laboratory of Computational Physics, Institute of Applied Physics and Computational Mathematics, Beijing 100088, China}

\author{Bo Sun}
\affiliation{Laboratory of Computational Physics, Institute of Applied Physics and Computational Mathematics, Beijing 100088, China}

\author{Hai-Feng Liu}
\affiliation{Laboratory of Computational Physics, Institute of Applied Physics and Computational Mathematics, Beijing 100088, China}

\author{Hai-Feng Song}
\affiliation{Laboratory of Computational Physics, Institute of Applied Physics and Computational Mathematics, Beijing 100088, China}

\thanks{Zi-Hang Chen and Jie Sheng contributed equally to this work.}
	
	\pacs{Zirconium hydride, Phase-field method, Temperature effect, Mismatch degree}
	\date{\today}
	
	\begin{abstract}
		Hydride precipitation in zirconium cladding materials can damage their integrity and durability. 
		Service temperature and material defects have a significant effect on the dynamic growth of hydrides.
		In this study, we have developed a phase field model based on the assumption of elastic behaviour within a specific temperature range (613-653K). This model allows us to study the influence of temperature and interfacial effects on the morphology, stress, and average growth rate of zirconium hydride.
		The results suggest that changes in temperature and interfacial energy influence the aspect ratio and average growth rate of the hydride morphology. The ultimate determinant of hydride orientation is the loss of interfacial coherence, primarily induced by interfacial dislocation defects and quantifiable by the mismatch degree $q$.
		An escalation in interfacial coherence loss leads to a transition of hydride growth from horizontal to vertical, accompanied by the onset of redirection behaviour. Interestingly, redirection occurs at a critical mismatch level, denoted $q_c$, and remains unaffected by variations in temperature and interfacial energy.
		However, this redirection leads to an increase in the maximum stress, which may influence the direction of hydride crack propagation.
	    This research highlights the importance of interfacial coherence and provides valuable insights into the morphology and growth kinetics of hydrides in zirconium alloys.

	\end{abstract}
	
	\maketitle
\renewcommand{\thefootnote}{\fnsymbol{footnote}}
\footnotetext[2]{$\,\,\,$These authors contributed eually to this work.}
\footnotetext[1]{$\,\,\,$Corresponding author.}

	\section{IINTRODUCTION}
	
	Zirconium (Zr) alloys are extensively employed as cladding materials in various reactor types\cite{hedayat2017developing} due to their low neutron absorption rate, robust resistance to corrosion, and adequate high temperature strength, excellent thermal conductivity, and ductility\cite{bair2015review,motta2019hydrogen}. However, hydrogen is absorbed into the Zr cladding by the waterside corrosion reaction during reactor operation\cite{ensor2017role}, thereby precipitating hydrides when the solid solubility limit of the hydrogen in the Zr alloys is exceeded. Hydride precipitates causes a degradation in the mechanical properties of the cladding\cite{ghosal2001corrosion}, increasing the embrittlement and even inducing the fracture of the cladding\cite{motta2019hydrogen,kim2022hydride}. Consequently, the investigation of the hydrogen corrosion behavior in Zr is crucial for predicting material property degradation and assessing the service life of the cladding.
	
	A great deal of experimental works have been done to better understand hydride precipitate and growth in Zr alloys\cite{kim2022hydride}. Based on experiments\cite{zuzek1990h,daum2009identification,puls2012effect}, there are four recognized phases of hydrides which can form in Zr, namely, $\zeta$-ZrH$_{0.5}$, $\gamma$-ZrH, $\delta$-ZrH$_{1.5+\rm{x}}$, and $\varepsilon$-ZrH$_2$ phases. $\delta$-phase is a stable face-centered cubic (fcc) phase and the most prevalent phase in the cladding materials. $\delta$-hydrides usually form on the circumferential planes of cladding in the absence of stresses, and have a platelet-like shape\cite{han2019phase,heo2019phase}. Moreover, when the cladding with hydrides is subjected to high temperature (350$^\circ$C for example) and applied tensile circumferential stress during dry storage, these hydride platelets undergo a transition from circumferential to radial orientations relative to the cladding tube geometry by dissolving and precipitating\cite{colas2010situ,colas2010hydride,colas2013effect}. The reoriented radial hydrides serve as fracture initiators and propagation, and ultimately cause the mechanical failure of the cladding. Therefore, the mechanical properties and performance of the Zr cladding are highly sensitive to the specific hydride microscopic morphology. However, more information is still needed in the areas of the formation mechanisms of various hydride morphology, particularly to develop the capability to simulate accurately these various morphology under varying conditions, such as different temperatures.
	
	Phase-field (PF) modeling, based on principles of thermodynamics and kinetics, provides a powerful approach for simulating microstructure evolution in various phase transformation phenomena, such as spinodal decomposition\cite{kwon2007coarsening,seol2003three}, coarsening\cite{kwon2009topology,mendoza2004topological}, solidification\cite{boettinger2002phase,YANG2020109220,yang2021hydride}, and film growth\cite{wang2004phase,SHI2022118147,shi2022ultrafast,shi2022phase}. Numerous PF models have been developed to investigate the microstructure evolution of different hydrides in Zr alloys. Most of them were created for the $\zeta$-hydride\cite{zhao2008characterization,thuinet2012phase,thuinet2013mesoscale} and the $\gamma$-hydride\cite{ma2002phase,ma2002simulation,ma2002effect,ma2006phase,shi2015quantitative,bair2016phase}. In recent years, many researchers have focused on the $\delta$-hydrides and started to develop corresponding PF models to study the characteristics of $\delta$-hydride morphology. Jokisaari \textit{et al.}\cite{jokisaari2015general} developed a $\delta$-hydride PF model to study the growth kinetics of the $\delta$-hydride by modifying the free energies based on the CALPHAD (CALculation of PHAse Diagrams). Bair \textit{et al.}\cite{bair2017formation} investigated the effects of metastable phases ($\zeta$-phase and $\gamma$-phase) on the nucleation and growth of the $\delta$-hydride using a multiphase field model. Han \textit{et al.}\cite{han2019phase} explored the stacking structure formation and transition of $\delta$-hydride precipitates in Zr. Heo \textit{et al.}\cite{heo2019phase} developed a polycrystalline PF model considering the loss of the interfacial coherency and the grain boundary segregation to investigate $\delta$-hydride growth behavior. Additionally, Lin \textit{et al.}\cite{lin2019modeling} and Shin \textit{et al.}\cite{shin2020phase} examined the reorientation behavior of $\delta$-hydrides under applied stress fields. Simon \textit{et al.}\cite{simon2021investigation} developed a quantitative PF model for $\delta$-hydrides and analyzed the role of anisotropic elastic interaction on hydride stacking and reorientation. Wu \textit{et al.}\cite{wu2022phase} studied the microstructural evolution of hydride blister formed by $\delta$-hydride aggregation. However, the morphology and growth kinetics of $\delta$-hydride are strongly influenced by the environmental temperature and metal/hydride interfacial effects (interfacial energy and interfacial coherency). Previous studies on $\delta$-hydrides in Zr lacked detailed discussions on the these key factors governing $\delta$-hydride morphology and growth kinetics, hindering the accurate simulation of $\delta$-hydride morphology in Zr and the development of quantitative PF model. Therefore, it is imperative to explore and comprehend the influences of these factors on $\delta$-hydride morphology and growth behavior in Zr.
	
	This work aims to investigate the influence of temperature and interfacial effects on the $\delta$-hydride morphology and growth kinetics. To achieve this objective, we adopt a PF model based on the \textit{Kim-Kim-Suzuki} (KKS) assumption\cite{kim1999phase}. Using CALPHAD method, temperature-dependent bulk free energy curves are constructed to describe the contribution of bulk free energy to the total free energy within the service temperature range of zirconium alloys. Therefore, we can capture the morphological changes of hydrides at different temperatures. In addition, the effects of interfacial energy and the loss of interfacial coherency on the morphology, stress and growth kinetics of hydrides are studied, and a reorientation behavior influenced by the loss of interfacial coherency is found. Meanwhile, we reveal the underlying mechanism of these factors affecting the hydride morphology and growth kinetics based on the diffusion interfacial velocity theory.
	
	The contents of the article are organized as follows. In Section 2, we presents this PF model coupled with varying temperature in detail, including the governing equations of the PF model and the free energies of the system and the interface moving velocity. In Section 3, the results and their implications are presented. Finally, Section 4 includes the paper by summarizing the key results and contributions of this study. 
	
	\section{PHASE-FIELD MODEL}
	
	\subsection{Governing equation}
	The PF method is widely used to study the evolution of microstructures with heterogeneity in composition and structure. Specifically, in the Zr-H system, the structural order parameter $\eta$ is used to describe the phase transformation from $\alpha\mbox{-}\textrm{Zr}$ matrix to $\delta\mbox{-}$hydride precipitate and the variation of hydrogen concentration $c$ is used to describe diffusion during the phase transformation. In this PF model, the Zr-H system's total free energy $F$ is consist of the bulk free energy $F_{\rm{bulk}}$, the interface gradient energy $F_{\rm{int}}$, and the elastic
	strain energy $F_{\rm{el}}$, which is given by:
	\begin{align}
		F=&F_{\rm{bulk}}+F_{\rm{int}}+F_{\rm{el}}\nonumber\\
		=&\int{\left[ f_{\rm{bulk}}(c,T,\eta)+ f_{\rm{int}}(\nabla\eta,\nabla c)+f_{\rm{el}}(\boldsymbol{u},\eta)\right] }\,dV,
	\end{align}
	where $f_{\rm{bulk}}(c,T,\eta)$ is the bulk free energy density as a function of temperature $T$, composition $c$ and structural order parameter $\eta$, $f_{\rm{int}}(\nabla\eta,\nabla c)$ is the interface gradient energy density because of diffuse interface, $f_{\rm{el}}(\boldsymbol{u},\eta)$ is the elastic strain energy density associated with displacement $\boldsymbol{u}$ and order parameter $\eta$.
	
	To ensure the reduction of the free energy during the evolution of microstructures, the PF governing equation could be derived by minimizing the total free energy $F$ via variational differentiation as follows:
	\begin{gather}
		\dfrac{\partial \eta }{\partial t}=-M\dfrac{\delta F}{\delta \eta },\\
		\dfrac{\partial c}{\partial t}=\nabla \cdot \left (D\nabla \dfrac{\delta F}{\delta c}  \right ),
	\end{gather}
	where $M$ is the kinetic coefficient; $t$ is the time; $D$ is the mobility similar to one defined by Steinbach and Apel as follows\cite{steinbach2006multi}: $D=D_0/(\partial^2 f_{\rm{bulk}}/\partial c^2$), where $D_0$ is the diffusion coefficient.
	
	\subsection{Bulk free energy density}
	
	The bulk free energy density can be constructed by KKS assumption. In the KKS assumption, each material point is regarded as a mixture of two phases, and a local equilibrium of chemical potential is always satisfied between the two phases. In this PF model, these assumptions could be expressed as follows:
	\begin{align}
		c = [1-h(\eta)]c_{\alpha} + h(\eta)c_{\delta},\\
		\frac{\partial f_{\alpha}(c_{\alpha},T)}{\partial c_{\alpha}} = \frac{\partial f_{\delta}(c_{\delta},T)}{\partial c_{\delta}},
	\end{align}
	where $c_{\alpha}$ and $c_{\delta }$ are the mole fraction of hydrogen in the $\alpha\mbox{-}$Zr matrix and $\delta\mbox{-}$hydride precipitate respectively; $f_{\alpha}(c_{\alpha},T)$ and $f_{\delta}(c_{\delta},T)$ are the free energy densities of the $\alpha\mbox{-}$Zr matrix and $\delta\mbox{-}$hydride precipitate at temperature $T$ respectively. Additionally, $h\left ( \eta  \right )=3\eta ^{2}-2\eta ^{3}$ is the interpolation function that increases monotonically from $h\left ( \eta =0 \right )=0$ to $h\left ( \eta =1 \right )=1$. The bulk free energy density $f_{\rm{bulk}}(c,T,\eta)$ of the system is defined by a method similar to the concentration assumption as follows:
	\begin{align}
		&f_{\rm{bulk}}(c,T,\eta)=h\left ( \eta  \right )f_{\delta }(c_{\delta},T)+\left [ 1- h\left ( \eta  \right )\right ]f_{\alpha }(c_{\alpha},T)+wg\left ( \eta  \right ),\\
		\label{equ6}
		&f_{\alpha }\left ( c_{\alpha},T \right )=\frac{G_{m}^{\alpha}}{V_m},\\
		\label{equ7}
		&f_{\delta }\left ( c_{\delta},T \right )=\frac{G_{m}^{\delta}}{V_m},
	\end{align}
	where $g\left ( \eta  \right )=\eta ^{2}\left ( 1-\eta ^{2} \right )$ is the double potential well function and $w$ is the height of the double potential well, $G_{m}^{\alpha}$ and $G_{m}^{\delta}$ are the molar Gibbs free energies of the $\alpha\mbox{-}$Zr matrix and $\delta\mbox{-}$hydride precipitate respectively; $V_m$ is the molar volume of the system.
	
	The molar Gibbs free energies $G_{m}^{\alpha}$ and $G_{m}^{\delta}$ can be obtained by CALPHAD method through the sublattice model\cite{dupin1999thermodynamic,zhong2012thermodynamics}. In the sublattice model, Zr fills the first sublattice and hydrogen and vacancies fill the second sublattice. Considering the $\alpha\mbox{-}\textrm{Zr}$ matrix with an hcp structure and $\delta\mbox{-}\textrm{hydride}$ precipitate  with an fcc structure, $\left (\rm{Zr}  \right )_{1}\left ( \rm{H},\Box  \right )_{1}$ is used to describe the $\alpha\mbox{-}\textrm{Zr}$ matrix and $\left (\rm{Zr}  \right )_{1}\left ( \rm{H},\Box \right )_{2}$ is used to describe $\delta\mbox{-}\textrm{hydride}$ precipitate. The molar free energies for $\alpha\mbox{-}\textrm{Zr}$ and $\delta\mbox{-}\textrm{hydride}$ is expressed by the following formula:
	\begin{gather}
		G_{m}^{\alpha}= G_{m}^{\rm{ref},\alpha}+G_{m}^{mix,\alpha},\\
		G_{m}^{\delta}= G_{m}^{\rm{ref},\delta}+G_{m}^{mix,\delta}+G_{m}^{ex,\delta},\\
		G_{m}^{\rm{ref},\alpha}=\frac{1}{1+y_{\rm{H}}}\left ( y_{\rm{H}}G_{\rm{Zr}:\rm{H}}^{0,\alpha}+y_{\Box }G_{\rm{Zr}:\Box }^{0,\alpha} \right ),\\
		G_{m}^{mix,\alpha}=\frac{1}{1+y_{\rm{H}}}RT\left ( y_{\rm{H}}lny_{H}+y_{\Box }lny_{\Box }\right ),\\
		G_{m}^{\rm{ref},\delta}=\frac{1}{1+2y_{\rm{H}}}\left ( y_{\rm{H}}G_{\rm{Zr}:\rm{H}}^{0,\delta}+y_{\Box }G_{\rm{Zr}:\Box }^{0,\delta} \right ),\\
		G_{m}^{mix,\delta}=\frac{2}{1+2y_{\rm{H}}}RT\left ( y_{\rm{H}}lny_{H}+y_{\Box }lny_{\Box }\right ),\\
		G_{m}^{ex,\delta}=\frac{1}{1+2y_{\rm{H}}}y_{\rm{H}}y_{\Box }\left [ L_{\rm{Zr}:\rm{H},\Box }^{0,\delta}+ L_{\rm{Zr}:\rm{H},\Box }^{1,\delta}\left ( y_{\rm{H}}-y_{\Box } \right )\right ],
	\end{gather}
	where $G_{m}^{\rm{ref},\alpha}$ and $G_{m}^{\rm{ref},\delta}$ are the reference free energy of mixing of the end members in the $\alpha\mbox{-}\textrm{Zr}$ matrix and $\delta\mbox{-}\textrm{hydride}$ precipitate respectively; $G_{m}^{mix,\alpha}$ and $G_{m}^{mix,\delta}$ are the free energy of mixing for an ideal soluation in the $\alpha\mbox{-}\textrm{Zr}$ matrix and $\delta\mbox{-}\textrm{hydride}$ precipitate respectively.
	Significantly, the free energy formula of $\delta\mbox{-}\textrm{hydride}$ precipitate compared with the matrix includes the excess free energy $G_{m}^{ex,\delta}$ of mixing. The parameters for  $G_{\rm{Zr}:\rm{H}}^{0,\alpha}$, $G_{\rm{Zr}:\Box}^{0,\alpha}$, $G_{\rm{Zr}:\rm{H}}^{0,\delta}$, $G_{\rm{Zr}:\Box}^{0,\delta}$, $L_{\rm{Zr}:\rm{H},\Box}^{0,\delta}$, $L_{\rm{Zr}:\rm{H},\Box}^{1,\delta}$ are supplied by Ref \cite{dupin1999thermodynamic,zhong2012thermodynamics}. $y_{\rm{H}}$ and $y_{\Box}$ are the mole fractions of hydrogen and vacancy atom related to $c_\alpha$ and $c_\delta$ respectively. These formulas for the $\alpha\mbox{-}\textrm{Zr}$ matrix can be expressed as follows:
	\begin{gather}
		y_{\rm{H}}=\frac{c_\alpha}{1-c_\alpha},\\
		y_{\Box}=1-\frac{c_\alpha}{1-c_\alpha}.
	\end{gather}
	These formulas for the $\delta\mbox{-}\textrm{hydride}$ precipitate can be expressed as follows:
	\begin{gather}
		y_{\rm{H}}=\frac{c_\delta}{2\left ( 1-c_\delta \right )},\\
		y_{\Box}=1-\frac{c_\delta}{2\left ( 1-c_\delta \right )}.
	\end{gather}
	
\begin{figure}[h]
		\centering
		\includegraphics[width=0.5\textwidth]{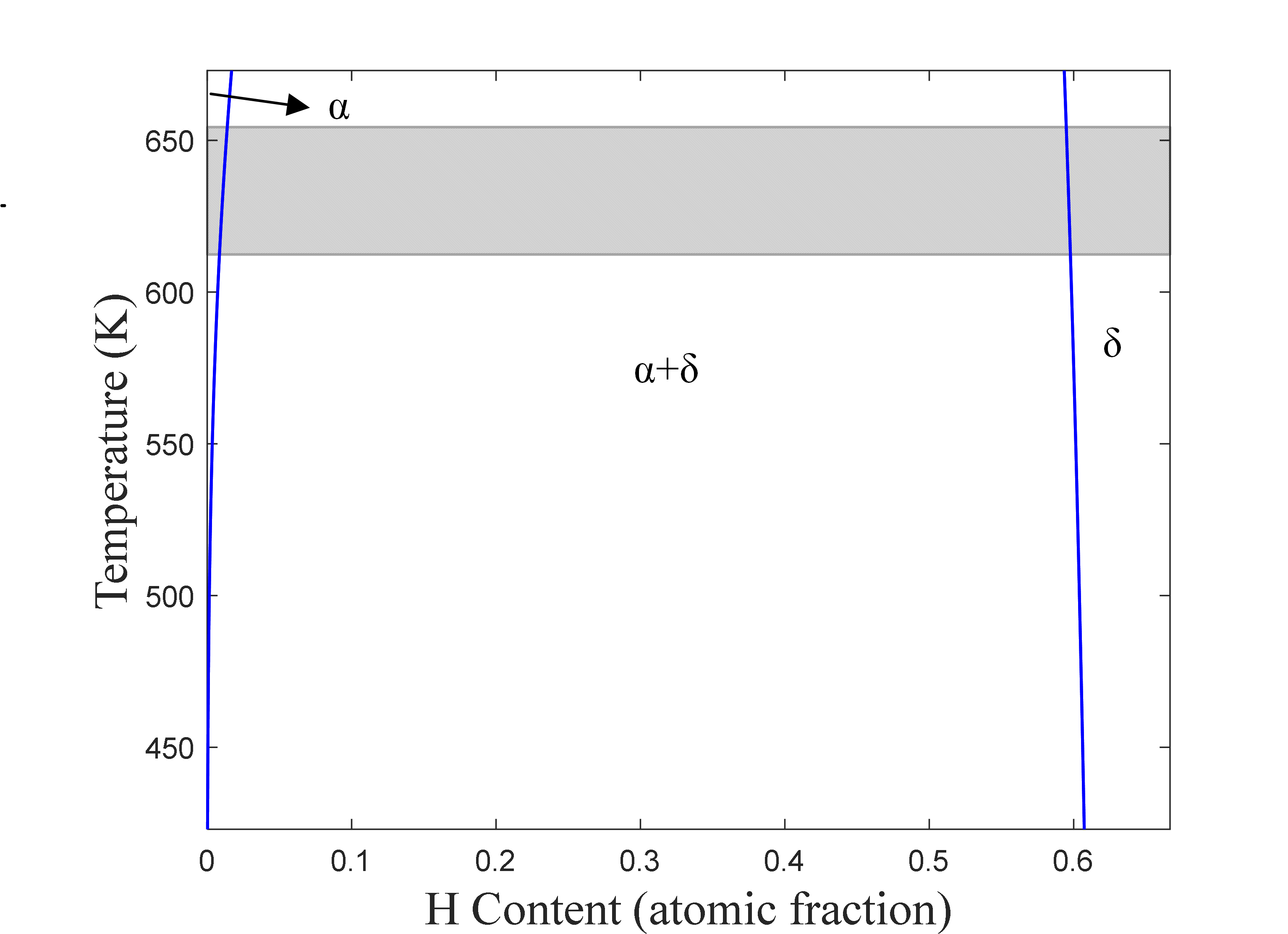}
		\caption{\label{fig1}Zr-H phase diagram. The simulation test in this work is conducted in accordance with the service temperature range of the Zr cladding tube, with 613-653 K (shaded part of the fig) selected as the operating temperatures.}
	\end{figure}

	According to the above CALPHA method, the Zr-H phase diagram is drawn in Fig. \ref{fig1}. To improve the numerical stability of the calculation, the molar Gibbs free energies $G_{m}^{\alpha}$ and $G_{m}^{\delta}$ are approximated by using quadratic polynomials near the equilibrium composition. Based on parabola approximation\cite{sheng2022phase,sheng2023multiphasefield,bair2017formation}, the molar Gibbs free energies $G_{m}^{\alpha}$ and $G_{m}^{\delta}$ are  written as:
	\begin{gather}
		G_{m}^{\alpha}=a_1\left (T  \right )c_{\alpha}^{2}+b_1\left (T  \right )c_{\alpha}+c_1\left (T  \right ),\\
		G_{m}^{\delta}=a_2\left (T  \right )c_{\delta}^{2}+b_2\left (T  \right )c_{\delta}+c_2\left (T  \right ),\\
		a_1\left (T  \right )=206.6\rm{e}^{\frac{4429}{\textit{T}}} \rm{[J/mol]},\\
		b_1\left (T  \right )=40.63T-8.918\times 10^{4} \rm{[J/mol]},\\
		c_1\left (T  \right )=-55.56T+6484 \rm{[J/mol]},\\
		a_1\left (T  \right )=-771.3T+8.305\times 10^{5} \rm{[J/mol]},\\
		b_1\left (T  \right )=1021T-1.109\times 10^{6} \rm{[J/mol]},\\
		c_1\left (T  \right )=-361.2T+3.216\times 10^{5} \rm{[J/mol]},
	\end{gather}
	where the detail determined method of coefficients can be seen in Secton 1 of Supplementary Materials. 
	
\begin{figure}[h]
		\vspace{-1.0cm}
		\centering
		\includegraphics[width=0.5\textwidth]{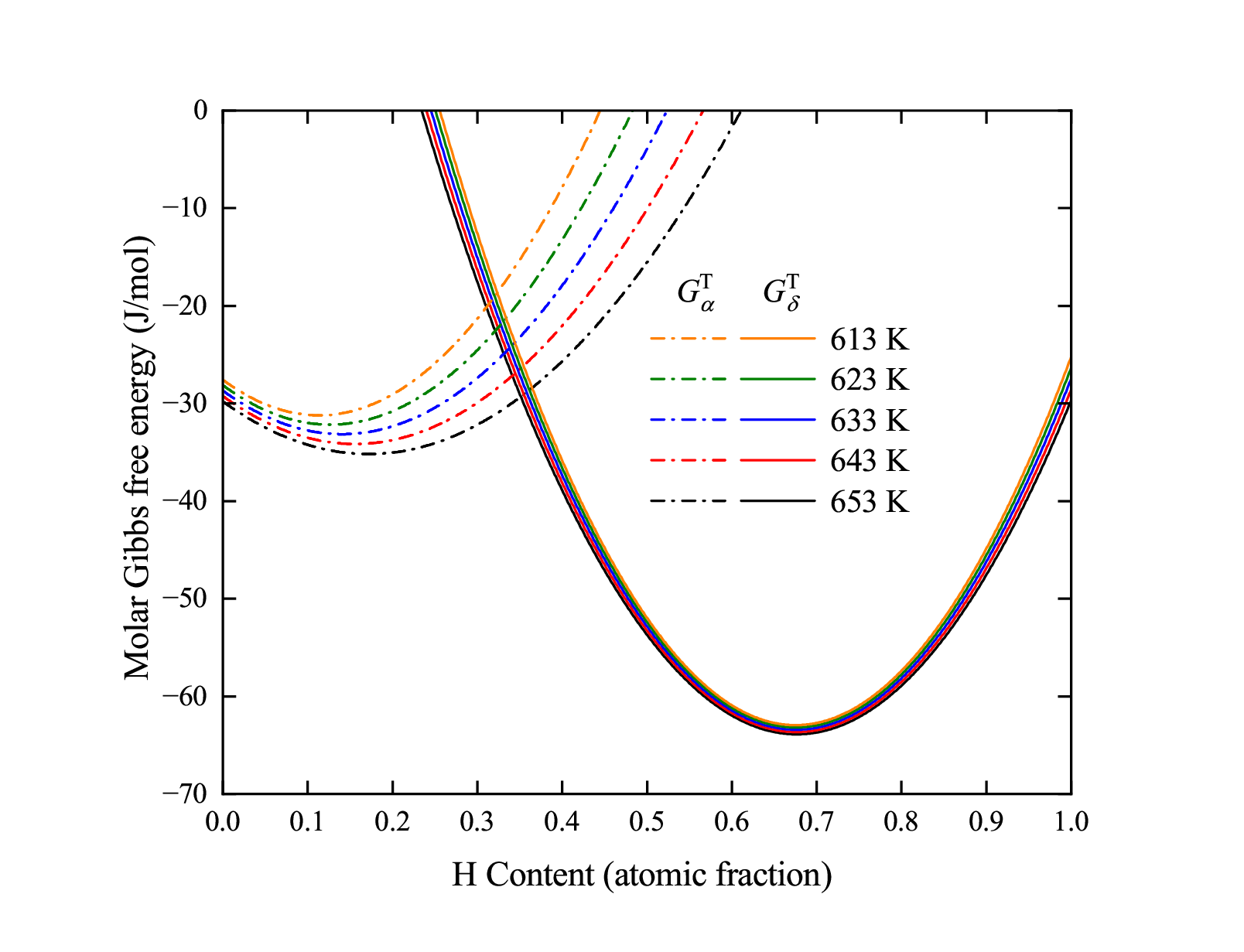}
		\caption{\label{fig2} Parabolic approximations for molar Gibbs free energies $G_{m}^{\alpha}$ and $G_{m}^{\delta}$ from T = 613 K to T = 653 K.}
	\end{figure}

	The service temperature of the Zr cladding tube is around 623 K. Fig. \ref{fig2} shows the results of the parabolic approximation for molar Gibbs free energies $G_{m}^{\alpha}$ and $G_{m}^{\delta}$ from $T$ = 613 K to $T$ = 653 K.
	
	\subsection{Interface gradient energy density}
	The interface gradient energy density includes two aspects of contribution from inhomogeneous composition and structure across the interface\cite{mai2016phase,han2019phase}. In the PF frame, the interface gradient density could be written as:
	\begin{equation}
		f_{\rm{int}}(\nabla\eta,\nabla c)=\frac{1}{2}k_{\eta }^{2}\left|\nabla\eta \right|^{2}+\frac{1}{2}k_{c}^{2}\left|\nabla c \right|^{2},
	\end{equation}
	where $k_{\eta}^{2}$ and $k_{c}^{2}$ are the gradient energy coefficients related to the order parameter gradient and the concentration gradient, respectively. Due to the fact that the contribution of order parameter gradient to the interface gradient energy could be sufficient to approximate the energy contribution from the diffuse interface, so it is assumed that $k_{c}^{2}=0$\cite{mai2016phase,han2019phase}. 
	
	The value of gradient energy coefficient $k_{\eta }^{2}$ and the height of the double potential well $w$ could be determined by interfacial energy $\sigma$ and interface thickness $\lambda$, as follows\cite{yang2021explicit}: 
	\begin{gather}
		k_{\eta }^{2}=\frac{3}{4}\sigma \lambda,\\
		w=24\frac{\sigma}{\lambda}.
	\end{gather}
	
	\subsection{Elastic strain energy density}
	Due to lattice mismatch and volume mismatch of two phases during the phase transformation from $\alpha\mbox{-}\textrm{Zr}$ matrix to $\delta\mbox{-}\textrm{hydride}$ precipitate, the elastic strain energy generated can not be ignored. The elastic strain energy density $f_{\rm{el}}$ is expressed as:
	\begin{equation}
		f_{\rm{el}}(\boldsymbol{u},\eta)=\frac{1}{2}C_{ijkl}\varepsilon_{ij}^{\rm{el}}\varepsilon_{kl}^{\rm{el}},
	\end{equation}
	where $\varepsilon_{ij}^{\rm{el}}$ is the elastic strain, $C_{ijkl}$ is the elastic constant tensor (the Einstein summation convention is used). Because solving the elastic solution of elastic inhomogeneous solid requires a lot of calculations, in our simulation, we assume that the elastic constant of the hydride phase is the same as that of the matrix\cite{han2019phase}. According to the classical Khachaturyan micro-elasticity theory\cite{morris2010khachaturyan}, the elastic strain can be calculated as follows:
	\begin{gather}
		\varepsilon_{ij}^{\rm{el}}=\delta\varepsilon_{ij}-\varepsilon_{ij}^{00}(q)h\left ( \eta  \right ),\\
		\delta\varepsilon _{ij}=\frac{1}{2}\left ( \frac{\partial u_{i}}{\partial x_{j}} +\frac{\partial u_{j}}{\partial x_{i}}\right ),
	\end{gather}
	where $\delta\varepsilon_{ij}$ is the microscopic heterogeneous strain relates to the displacement $\boldsymbol{u}$. $\varepsilon_{ij}^{00}(q)$ is the eigenstrain related to the mismatch degree $q$, depicting the degree of lattice mismatch due to the volume expansion during the phase transformation and the loss of interfacial coherency. The mismatch degree $q$ describes the loss degree of interfacial coherency. Notably, only elastic strain is taken into account in this PF model, which means that no plastic deformation could take place.
	
	With Lagrangian finite-strain theory\cite{lubliner2008plasticity}, the eigenstrain is expressed as:
	\begin{align}
		&\varepsilon_{ij}^{00}(q)=\frac{1}{2} \left ( F^{T}F-I \right )\\
		&F=\left (\dfrac{1}{q}  \right )^{\dfrac{1}{3}}\cdot\begin{bmatrix}
			\dfrac{\dfrac{\sqrt{2}}{2}a_{\delta } }{a_{\alpha }} & 0 &0 \\ 
			0 & q\cdot\dfrac{\dfrac{2\sqrt{3}}{3}a_{\delta } }{c_{\alpha }} &0 \\ 
			0&  0& \dfrac{\dfrac{\sqrt{2}}{2}a_{\delta } }{a_{\alpha }}
		\end{bmatrix}
	\end{align}
	where $F$ is the total lattice transformation tensor or deformation gradient tensor\cite{heo2019phase}, the superscript $T$ represents the matrix transpose and $I$ represents the identity matrix, $a_ {\alpha} $, $c_ {\alpha} $ and $a_ {\delta} $ represents the lattice parameters of $\alpha\mbox{-}$Zr and $\delta\mbox{-}\textrm{hydride}$. The calculation process of $F$ is detailed in Section 2 of Supplementary Materials. 
	
	Once the elastic constant tensor and strain are defined, the stress $\sigma_{ij}$ assuming linear elasticity is derived as follows: 
	\begin{equation}
		\sigma _{ij}= C_{ijkl}\varepsilon_{kl}^{\rm{el}}.
	\end{equation}
	The displacement of system is obtained by assuming the use of quasi steady state approximation to solve the mechanical equilibrium equation, as follows
	\begin{equation}
		\nabla\cdot \sigma _{ij}=0.
	\end{equation}
	
	\begin{table*}[ht]
		\caption{\label{tab1}Model parameters used in the simulations.}
		\renewcommand{\arraystretch}{1.5}
		\begin{tabular}{cccc}
			\hline
			Parameters                                 & Symbol     & Value           & Unit\\
			\hline
			Interface thickness                        & $\lambda$  & $4\times 10^{-9}$            & $\rm m$ \\                           
			Diffusion coefficient\cite{han2019phase}   & $D_0$      & $7\times 10^{-7}{\rm e}^{(-44560/RT)}$     & $\rm m^2/s$	\\
			Gas constant                               & $R$        & 8.3145          & $\rm J/\left ( mol\cdot K \right )$ \\
			Molar volume\cite{jokisaari2015general}    & $V_m$      & 14              & $\rm cm^3/mol$ \\
			Kinetic coefficients\cite{han2019phase}    & $M$        & $5\times 10^{-4}$  & $\rm m^3/\left ( J\cdot s \right )$ \\
			Elastic constant\cite{olsson2014ab}        & $C_{11},C_{12},C_{13},C_{33},C_{44}$     & 155, 67, 65, 155, 40    & $\rm GPa$ \\
			Eigenstrain\cite{han2019phase}             & $\varepsilon _{ij}^{00}$     & $\dfrac{1}{2}\left (\left (\dfrac{1}{q}  \right )^{\dfrac{2}{3}}\cdot
			\begin{bmatrix}
				1.07751 & 0 &0 \\ 
				0 & \left(\dfrac{1}{6}+\dfrac{4}{3}q^2\right)\cdot0.86821 &0 \\ 
				0&  0& 1.07751
			\end{bmatrix}-I\right )$      & - \\
			Misfit parameter\cite{heo2019phase}        & $q$        & 0.8-1.0         & - \\
			Temperature                                & $T$        & 613-653         & K \\
			Interfacial energy\cite{han2019phase}      & $\sigma$   & 0.28-1.0        & $\rm J/m^2$ \\
			\hline
		\end{tabular}	
	\end{table*}
	
	\subsection{Interface moving velocity} 
	In the classical sharp interface kinetics, the normal moving velocity of the interface is proportional to the variation of free energy per unit volume during the nucleation and growth process\cite{porter2009phase}. For the PF diffuse interface kinetics, Yang \textit{et al.}\cite{yang2021explicit} adopted the operator analysis method to analyze the PF kinetics equation, and derived the expression of the interface moving velocity. The specific expression is as follows:
	\begin{gather}
		v_{\textrm{int}}=\dfrac{3}{4}M\lambda \left (\Delta G -\sigma H_{\textrm{curv}} \right )\label{eq30},\\
		\Delta G =\Delta G_{\rm chem}-\Delta G_{\rm{el}},\\
		\Delta G_{\rm chem}=f_{\alpha }\left ( c_{\alpha},T \right )-f_{\delta }\left ( c_{\delta},T \right )-\dfrac{\partial f_{\alpha }\left ( c_{\alpha},T \right ) }{\partial c_{\alpha}}\left ( c_{\alpha}-c_{\delta} \right ),\\  
		\Delta G_{\rm{el}}=-\sigma _{ij}\varepsilon _{ij}^{00}(q)\label{eq32},
	\end{gather}
	where $R_{\textrm{curv}}$ represents the local curvature, $\Delta G$ represents the change of free energy during the phase transformation, including two parts: chemical driving force $\Delta G_{\textrm{chem}}$ and elastic interaction energy $\Delta G_{\rm{el}}$. It can be seen from the above formula that the interface moving velocity is mainly affected by the dynamic coefficient, interfacial energy, chemical driving force, elastic interaction energy and interface curvature. In Eq. \ref{eq30}, the first term promotes the occurrence of the movement of the interface, and the second term inhibits the advancement of the interface, and the smaller the radius of curvature, the more obvious the inhibition effect. When the elastic interaction energy ($\Delta G_{\rm{el}}$) is negative, it exerts a positive influence on interface pushing.  Conversely, when $\Delta G_{\rm{el}}$ is positive, it exhibits an inhibitory effect on interface pushing\cite{han2019phase}.
	
	\subsection{Numerical implementation}
	In this study, the finite element method (FEM)\cite{bair2016phase,sheng2022phase} is used to solve the governing equation. To ensure both stability and accuracy in the numerical simulations, a triangular element with a grid size of 2 nm is chosen. Additionally, a local adaptive mesh refinement scheme is applied strategically to avoid unnecessary refinement of finite element meshes throughout the simulation domain, thus enhancing computational efficiency.
	
	For time integration, we adopt the implicit backward difference method, which provides numerical stability when solving the time-dependent governing equation. To further optimize the time stepping process, an adaptive time marching scheme is implemented with a tolerance level set at $10^{-3}$. This adaptive approach allowed for efficient time stepping while still maintaining the required level of accuracy for the simulation results. 
	
	\begin{figure*}[ht]
		\centering
		\includegraphics[width=1.0\textwidth]{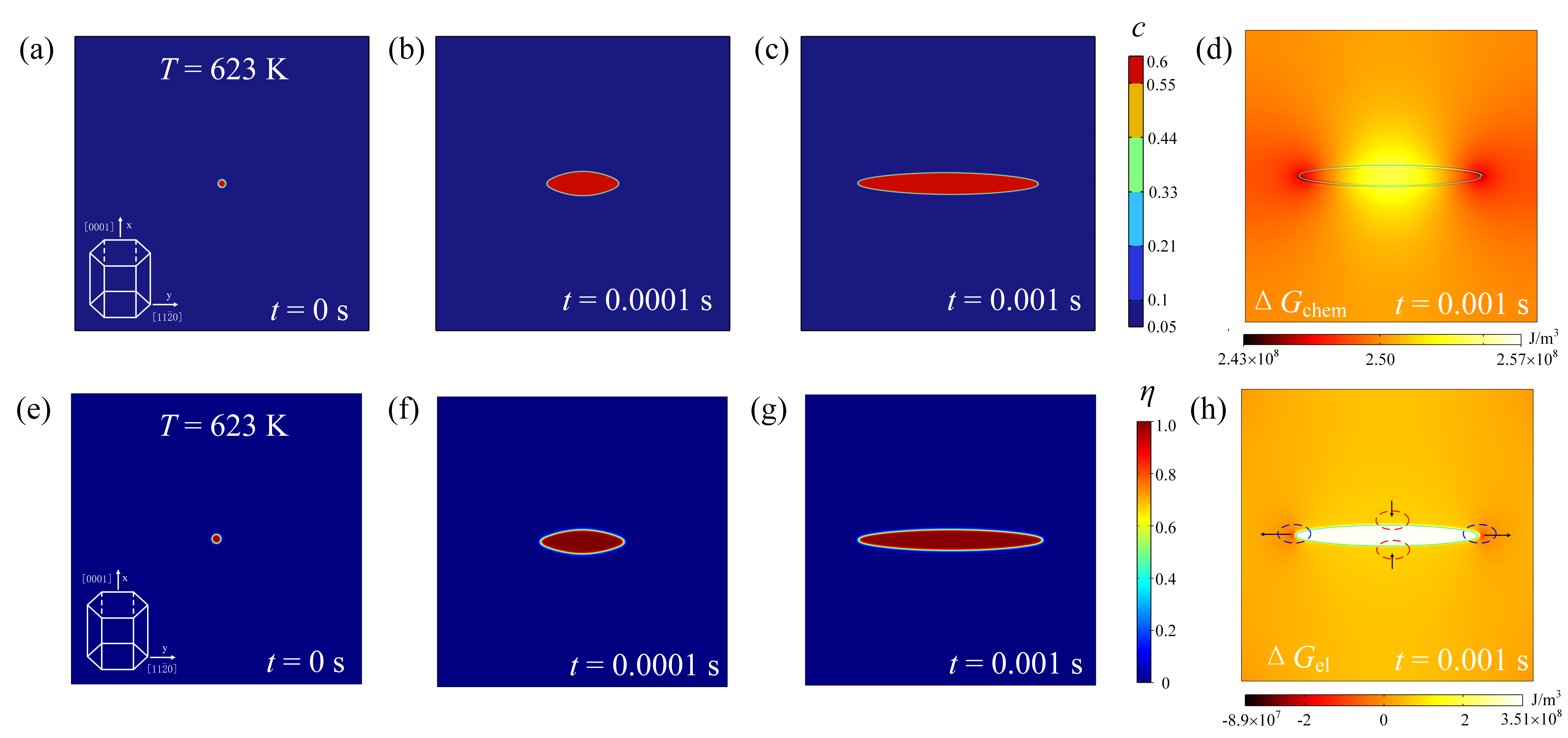}
		\caption{\label{fig3} Evolution of $c$ (a), (b), (c) and $\eta$ (e), (f), (g) as a function of time for systems, a single h.c.p Zr crystal describes the crystalline orientation of the system(The mismatch degree q is equal to 1); (d) Distribution Diagram of chemical driving force of $\delta\mbox{-}\textrm{hydride}$. ; (h) Distribution Diagram of elastic interaction energy of $\delta\mbox{-}\textrm{hydride}$.}
	\end{figure*}	
	
	\section{RESULTS AND DISCUSSION} \label{sec3} 
	The initial hydrogen concentration in the matrix is set to 0.05, while the hydride has an initial concentration of 0.6 \cite{han2019phase}. Nucleation of the hydride occurres at the center of the simulation domain, with a nucleation point having a radius of 2 nm. The presence of hydrogen in the matrix at a supersaturated level serves as a larger driving force for hydride growth\cite{han2019phase}, ultimately reaching a two-phase equilibrium. Table \ref{tab1} provides an overview of all the parameters utilized in this simulation. 
	
	To validate the effectiveness of the model, we initially conducted tests under completely coherent conditions with a mismatch degree of $q$ = 1, consistent with previous results. These tests confirmed the reliability and accuracy of our model in capturing the growth morphology of hydrides. As illustrated in Fig. \ref{fig3} (a) - (c) and (e) - (g), the phase transition from the metal matrix to hydride precipitation has been effectively monitored and characterized using variables $c$ (concentration) and $\eta$ (order parameter). In Fig. \ref{fig3} (a) and (e), the system orientation is specified such that the x-axis (horizontal direction) and the y-axis (vertical direction) align with the a-axis ($\left \{11\bar{2}0\right \}_\alpha$ direction) and c-axis ($\left \{0001\right \}_\alpha$ direction) of a hcp Zr crystal. Due to the mismatch of the hydride phase being greater along the a-axis direction than along the c-axis direction, hydride growth predominantly occurs in the a-axis direction, resulting in an elliptical shape of the hydride particles.
		
	\begin{figure*}[ht]
		\centering
		\includegraphics[width=1.0\textwidth]{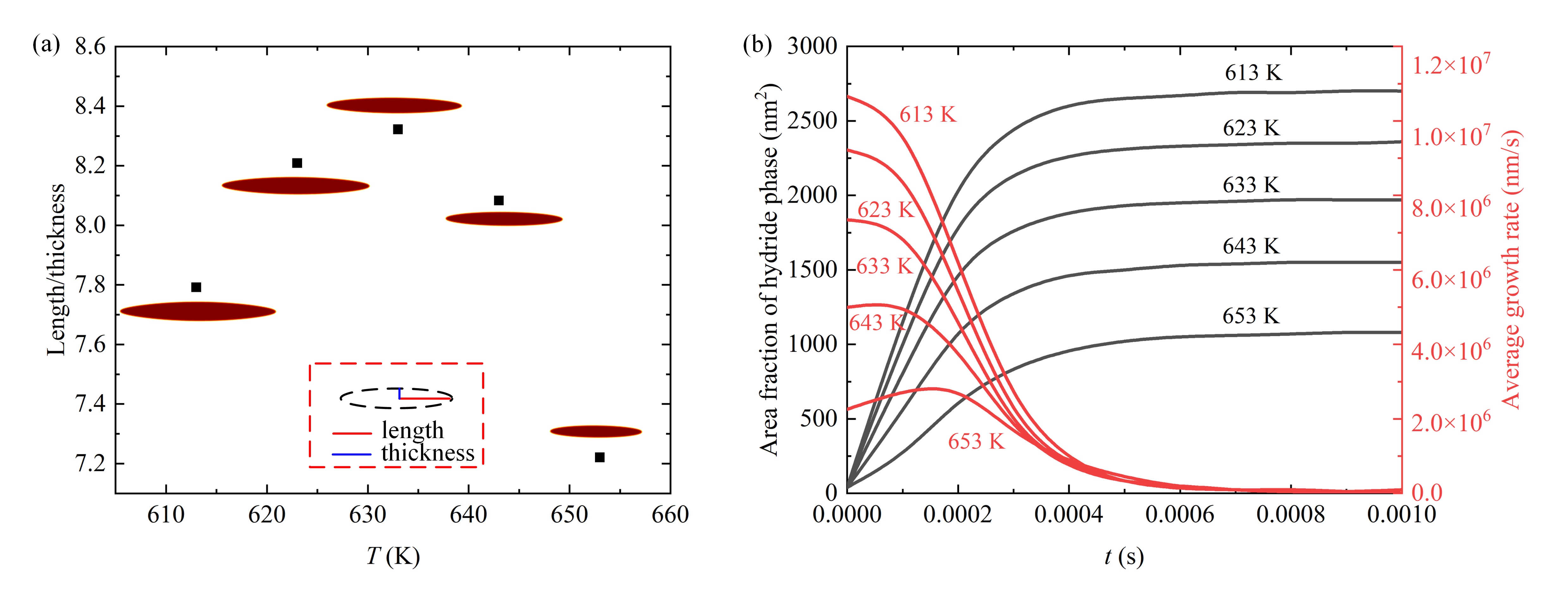}
		\caption{\label{fig4}(a) The length/thickness ratio of  $\delta\mbox{-}\textrm{hydride}$ after 0.001 s evolution at different temperatures; (b) Area fraction of hydride phase and average growth rate at different temperatures.}
	\end{figure*}
	The energy distribution analysis of the hydride, illustrated in Fig. \ref{fig3}(b), reveals a relatively uniform distribution of the chemical driving force, with slightly lower values observed at the hydride tip. In contrast, the distribution of elastic interaction energy exhibits greater anisotropy compared to the chemical driving force. At the hydride tip, the negative elastic interaction energy favors hydride growth along the a-axis. The positive elastic interaction energy can inhibit hydride growth along the c-axis. These results highlight the important role of elastic interaction energy in shaping hydride morphology. Next, the effects of temperature, interfacial energy and mismatch degree on the kinetic growth of hydride were tested.
    	
    \begin{figure*}[ht]
    	\centering
    	\includegraphics[width=1.0\textwidth]{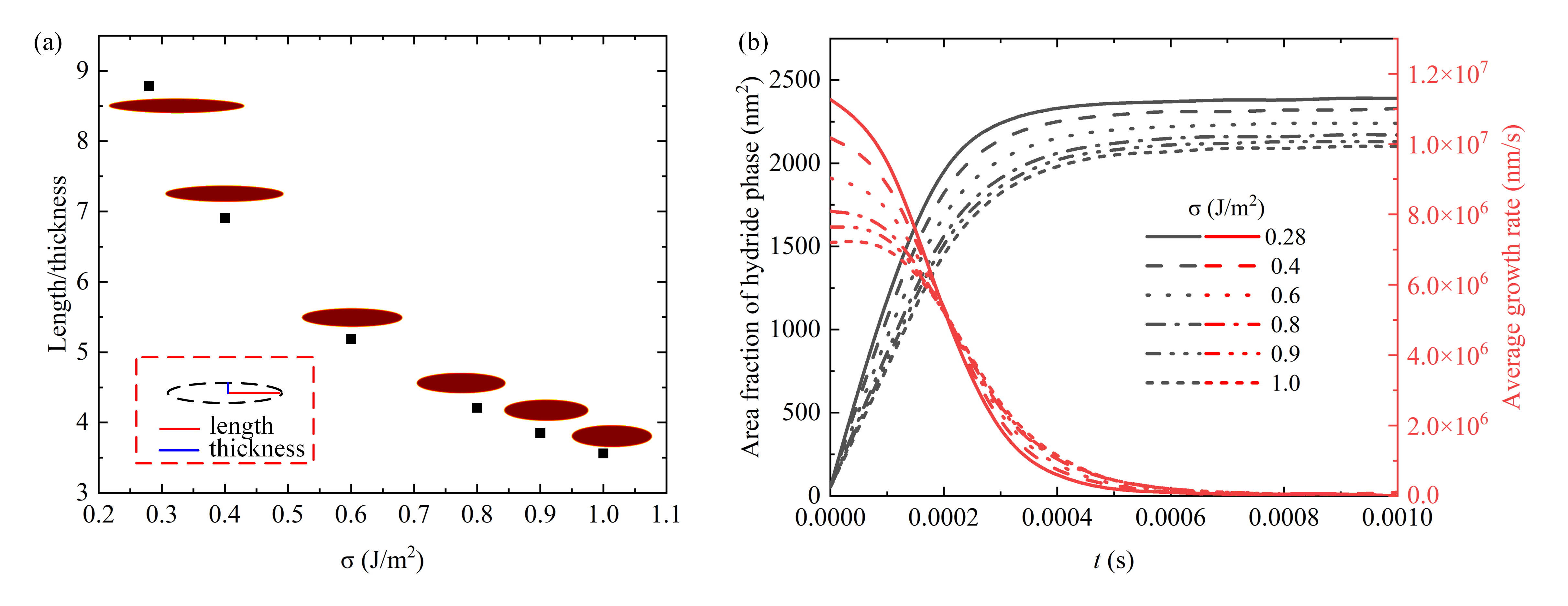}
    	\caption{\label{fig5}(a) The length/thickness ratio of  $\delta\mbox{-}\textrm{hydride}$ after 0.001 s evolution at different interfacial energies; (b) Area fraction of hydride phase and average growth rate at different interfacial energies.}
    \end{figure*}
	
	\subsection{Effect of temperature}
	Under service conditions at approximately 623 K, hydrogen can exist in solution where it is highly mobile. However, the presence of excess hydrogen or changes in temperature may lead to the precipitate of hydrides \cite{steinbruck2010high}. In order to investigate the effect of temperature on $\delta$-hydride, simulations were performed with consistent initial conditions at temperatures of 613 K, 623 K, 633 K, 643 K, and 653 K. The focus was on examining the influence of temperature on hydride shape and size.

	Within this temperature range, the diffusion coefficient exhibits variations as a function of temperature, while no significant changes are observed in the elastic constants and eigenstrain. It is assumed that the interfacial energy remains relatively constant. For these reasons, all simulations were conducted using the elastic constants, constitutive strains, and temperature-dependent diffusion coefficients as presented in Table 1.
	
	The simulation results presented in Fig. \ref{fig4}(a) demonstrate the significant impact of temperature on the length and thickness of hydrides. Specifically, the length-to-thickness ratio of hydrides initially increases and then decreases as the temperature rises. The variation in precipitate thickness is likely attributed to changes in the chemical free energy contributions within the system. Under equilibrium conditions, the shape of the precipitate is governed by the interplay between elastic properties and interfacial energy. The observed changes in hydride length and thickness can be primarily attributed to alterations in the chemical free energy resulting from temperature changes.
	
	To further investigate this phenomenon, the evolution of the growth area over time at different temperatures and the average growth rate were explored, as depicted in Fig. \ref{fig4}(b). As the temperature increases, the growth area of hydrides decreases, accompanied by a decrease in the growth rate. This decrease in the growth rate can be attributed to the reduction in the chemical driving force resulting from the increased temperature. According to Equation (\ref{eq30}), a decrease in the chemical driving force impedes the movement of the interface during the phase transition. It is important to note that when testing a wider temperature range, the results did not converge. Considering the significant temperature variations, the competition between elastic properties and interfacial energy governs the shape of the precipitate under equilibrium conditions. Therefore, it is likely that adjustments to the elastic strain energy or interfacial energy are required to achieve convergence when considering temperature changes.
	\begin{figure*}[htb]
		\centering
		\includegraphics[width=1.0\linewidth]{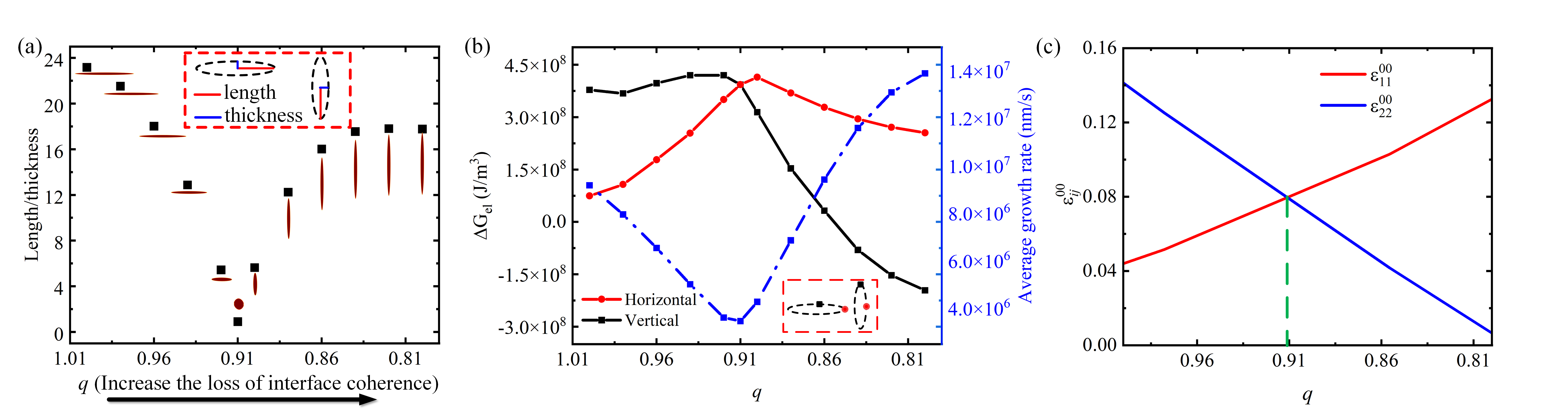}
		\caption{\label{fig6}(a) The morphology and length-to-thickness ratio of hydrides under varying mismatch conditions; (b) The average growth rate in both horizontal and vertical directions, along with the corresponding elastic interaction energy (represented by the red dot in the horizontal direction and the black dot in the vertical direction at the interface at time $t$=0.0002s) at different mismatch levels; (c) Eigenstrain at various mismatch levels.}
	\end{figure*}
	
	\subsection{Effect of interfacial energy}
	
	The loss of interfacial coherency between the hydride and matrix has a significant effect on the interfacial energy between the $\alpha$ and $\delta$ phases. Jokisaari et al. \cite{jokisaari2016multiphysics} suggest that the interfacial energy can be estimated from the chemical energy of the interface and the energy contributed by the existence of dislocations, resulting in a combined energy. Experimental and theoretical calculations have yielded different estimates for the $\alpha$-$\delta$ interfacial energy, with Jokisaari et al. reporting approximately 0.3 $\rm{J/m^2}$, and Massih et al. \cite{massih2009stress} estimating it to be 0.28 $\rm{J/m^2}$ from the measurement data of nucleation rate. However, the excess energy caused by semi-coherency cannot be ignored. For convenience, this study considers an interfacial energy of 0.28 $\rm{J/m^2}$ and investigates the impact of interfacial effects with different interfacial energy values on the length of hydrides. It should be noted that the isotropy assumption of the interface is made due to the lack of data in the literature, but the orientation relationship between different directions may affect the interfacial energy value.
	
	As depicted in Fig. \ref{fig5}(a), the interfacial energy influences the morphology of the hydride particles but does not affect the equilibrium hydrogen composition in the two phases. Higher values of interfacial energy lead to the formation of spherical hydrides rather than the elongated sheet-like hydrides observed in experiments. The growth area changes over time under different interfacial energy conditions are illustrated in Fig. \ref{fig5}(b), indicating that the growth area is larger under low interfacial energy. In terms of the growth rate, higher interfacial energy results in a slower growth rate of hydrides, and hydrides with lower interfacial energy reach equilibrium faster. The interfacial energy affects the gradient energy coefficient in the phase field model and hinders the interface motion, which is consistent with the simulation results based on Equation (\ref{eq30}). Due to the difficulty in directly measuring the interfacial energy in experiments, theoretical calculations and the observed hydride morphology often align well with the calculated value of 0.28 $\rm{J/m^2}$. Generally, the interfacial energy can encompass both structural and coherent contributions due to the presence of dislocations at the interface. Considering the uncertainties in the simulation process, it is reasonable to appropriately increase the interfacial energy to obtain the ideal hydride morphology.
     	
     \begin{figure*}[ht]
     	\centering
     	\includegraphics[width=1.0\textwidth]{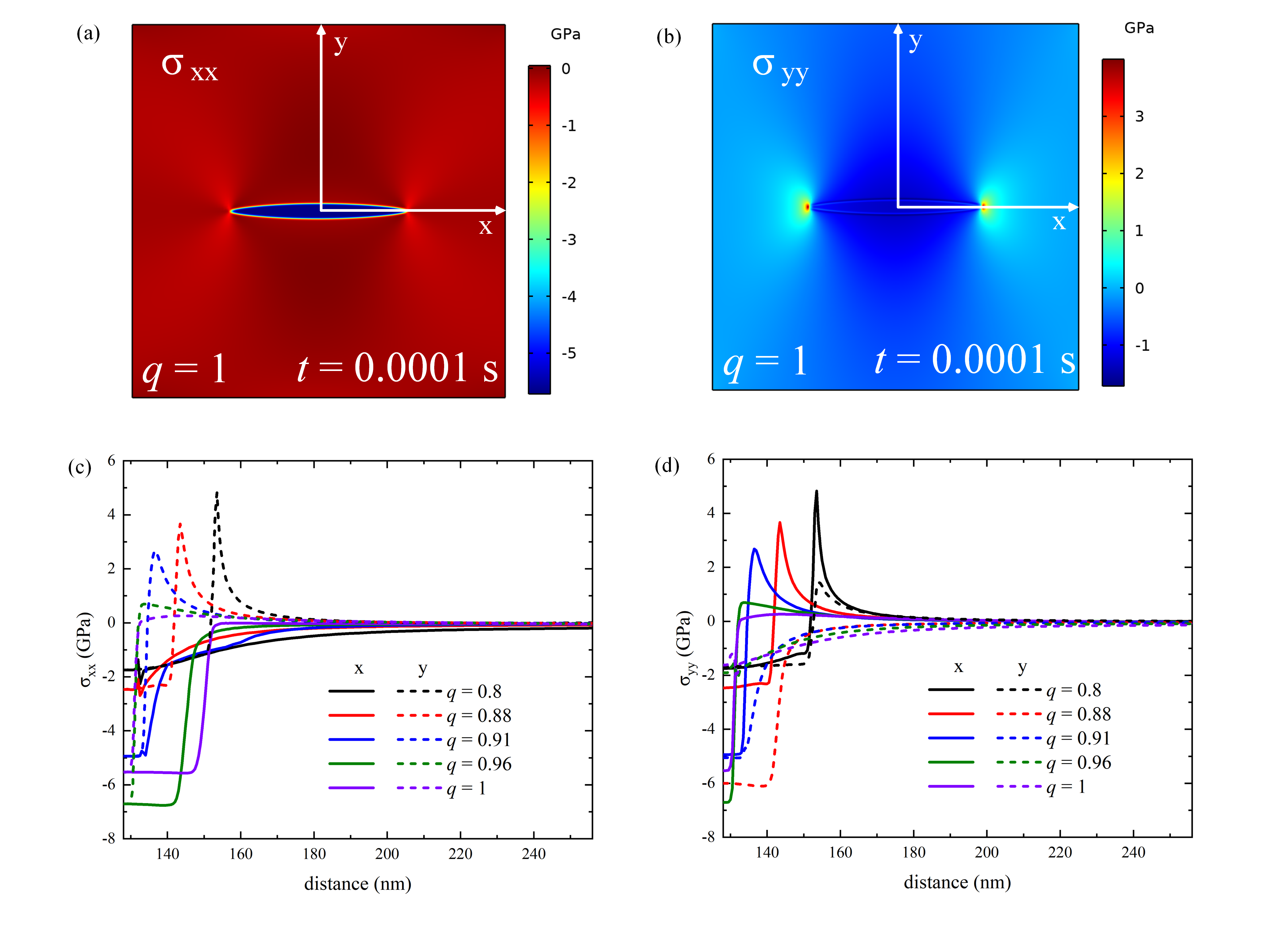}
     	\caption{\label{fig7}(a) Distribution of principal stress $\sigma_{xx}$; (b) Distribution of principal stress $\sigma_{yy}$; (c) $\sigma_{xx}$ curves along the x and y directions under different mismatch values; (d) $\sigma_{yy}$ curves along the x and y directions under different mismatch values.}
     \end{figure*}
	
	\subsection{Effect of the loss of interfacial coherency}
		
	\begin{figure*}[ht]
		\centering
		\includegraphics[width=1.0\textwidth]{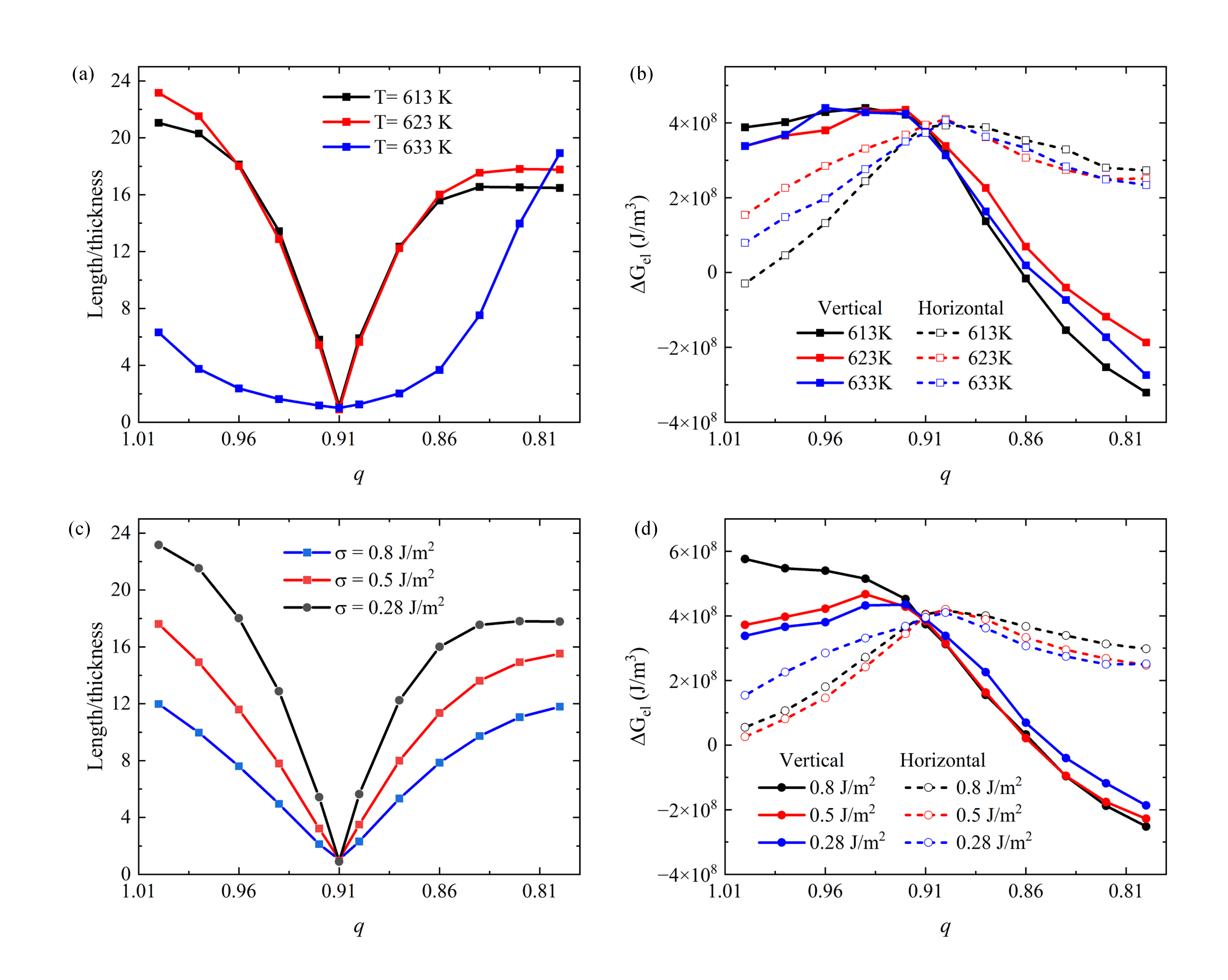}
		\caption{\label{fig8}(a) The aspect ratio of hydrides at different temperature and mismatch levels; (b) Elastic interaction energy in horizontal and vertical directions at different temperatures and mismatches; (c) The aspect ratio of hydrides at different interfacial energy and mismatch levels; (d) Elastic interaction energy in horizontal and vertical directions at different interfacial energy and mismatches.}
	\end{figure*}
	When considering interface effects, the influence of the loss of interfacial coherency should not be overlooked in addition to the interfacial energy. Liu \textit{et al.}\cite{liu2022dislocation} underscored the significant role of dislocations in hydride precipitation in their experiments. Dislocations can impact the lattice matching between two phases, leading to a loss of interfacial coherence. This loss of interfacial coherence, coupled with changes in interfacial energy, plays a pivotal role in shaping interfacial behavior. Due to the different crystal structures between the Zr-based phase and the hydride, the elastic strain energy generated by the precipitate of hydrides in the Zr-based phase cannot be ignored. The formation of hydrides involves significant volume expansion, which typically leads to the loss of interface coherency due to the formation of dislocations that accommodate the volume change. Different degrees of interface coherency loss result in varying degrees of lattice mismatch between the two phases. Here, by considering the mismatch degree introduced by Heo \textit{et al.}\cite{heo2019phase}, which acts on the transformation of the matrix phase to the hydride phase, the variation of mismatch degree primarily affects the magnitude of the eigenstrain. The calculation process for the mismatch degree can be found in Section 2 of Supplementary Materials. Nonetheless, current research has omitted the consideration of shear strain, as the path with the lowest energy for hydride phase transition is devoid of shear strain\textit{et al.}\cite{zhang2016homogeneous,louchez2017microscopic}, a similar method also adopted by Simon\cite{simon2021investigation}.
	
	The morphology of hydrides was simulated within the range of mismatch degrees from 0.8 (incompletely coherent interface) to 1 (fully coherent interface), as shown in Fig. \ref{fig6}(a). The aspect ratio of hydrides initially exhibits a decrease, followed by an increase as interfacial coherence loss escalates. When $q$ exceeds $0.91$, hydrides undergo redirection behavior and we designate this threshold as the critical mismatch degree $q_c$. This phenomenon can be elucidated by considering the anisotropic effect induced by the loss of interfacial coherence in the distribution of elastic interaction energy within hydrides. As depicted in Fig. \ref{fig6}(b), with a decreasing $q$ value, the elastic interaction energy initially exhibits an increase, followed by a decrease. When $q$ exceeds $q_c$, the elastic interaction energy in both horizontal and vertical directions is positive, which hinders hydride growth. Nevertheless, the elastic interaction energy in the horizontal direction is lower than that in the vertical direction, indicating that the inhibitory effect in the horizontal direction is comparatively less pronounced than in the vertical direction. Consequently, hydrides tend to grow horizontally. Conversely, when $q$is below $q_c$, the horizontal elastic interaction energy exceeds the vertical elastic interaction force, while the vertical elastic interaction energy exhibits a negative value when $q$ is below 0.86, which favors the growth of hydrides. Additionally, as the degree of mismatch increases, the average growth rate initially increases and then slows down, reaching its maximum value at $q_c$, corresponding to the minimum aspect ratio of hydride morphology. The average growth rate of hydrides decreases with an increase in elastic interaction energy, reflecting the inhibitory effect of elastic interaction energy on the hydride growth rate, consistent with the inhibition of interface migration velocity by elastic interaction energy in Equation (\ref{eq30}). From Fig. \ref{fig6}(c) , we can visually observe the changes in the principal strain as the mismatch parameter decreases. When the $q$ value is greater than $q_c$, the strain in the 22 direction is larger than that in the 11 direction. However, when the $q$ value is less than $q_c$, the strain in the 22 direction becomes smaller than that in the 11 direction. When the strain in the 11 direction exceeds that in the 22 direction, hydrides exhibit a propensity to grow in the 11 direction. This provides a model-based explanation for the occurrence of reorientation and the above analysis can provide trends of morphological evolution for the growing phase before reaching equilibrium.

	From the perspective of stress distribution, the principal stresses $\sigma_{xx}$ and $\sigma_{yy}$ exhibit negative values within the interior of the hydrogen precipitates, reaching maximum values at the tips in Fig. \ref{fig7}(a) and (b). As depicted in Fig. \ref{fig7}(c) and (d), a reduction in the mismatch value results in an increase in the maximum stress value. Initially, $\sigma_{xx}$ increases along the x-direction with an increase in distance and subsequently stabilizes, while $\sigma_{xx}$ follows an initial increase along the y-direction and subsequently decreases with distance. In contrast to $\sigma_{xx}$, the $\sigma_{yy}$ curve is reversed, which is attributed to the redirection phenomenon. In general, redirection induces alterations in stress and corresponding stress maxima. Through our stress analysis, we posit that redirection behavior, stemming from the loss of interfacial coherence, may lead to changes in the crack propagation direction at the hydride tip.

	To further examine the impact of temperature and interfacial energy on the mismatch degree in interface effects, we investigate the changes in the aspect ratio and elastic interaction energy distribution of hydrides at different mismatch degrees and temperatures (613K, 623K, and 633K). Additionally, we analyze the aspect ratio and elastic interaction energy distribution of hydrides under an interfacial energy value of 0.8 $\rm{J/m^2}$. The results from Fig. \ref{fig8}(a) and (b) illustrate that temperature influences the length-to-diameter ratio of hydrides at various mismatch degrees, while the overall trend of the length-to-diameter ratio remains consistent. Notably, the smallest aspect ratio of hydrides is consistently observed at a mismatch degree of $q_c$. Furthermore, irrespective of temperature variations, the change trend in elastic interaction energy remains consistent. At $q_c$, the elastic interaction energy values in the two directions become the closest, resulting in the formation of round-shaped hydrides. Similarly, the analysis from Fig. \ref{fig8}(c) and (d)  reveals that interfacial energy does not significantly impact the redirection of hydrides or the distribution trend of elastic interaction energy at $q_c$.
	
    In summary, our research encompasses a comprehensive investigation of hydrides, incorporating the concept of interfacial coherence loss. This leads to the identification of the phenomenon of hydride redirection, primarily attributed to alterations in the distribution of elastic energy. Redirection behavior is found to be associated with stress reversal, and within the interplay of temperature, interfacial energy, and interfacial coherence, the latter emerges as a prominent influencing factor. These insights provide valuable contributions to our understanding of how interfacial coherence loss impacts the morphology and behavior of Zr hydrides.
	
	\section{CONSLUSION} \label{sec4}
	
	In this study, the phase-field (PF) method was used to investigate the influence of temperature and interfacial effects on the kinetic process of hydride formation. By constructing two-phase free energy curves at different temperatures using the CALPHAD thermodynamic approach, the following main results were observed:
	
	1) Within a specific temperature range, temperature and interface conditions can indeed impact the length-to-thickness ratio and growth rate, while they do not appear to exert any influence on the redirection behavior. The aspect ratio of hydrides first increase and then decrease as the temperature rises. The increase of interfacial energy results in the decrease of the length-thickness ratio of the hydride. Additionally, the growth rate of hydrides demonstrates a linear decrease with an increase in temperature and interfacial energy.
 
	2) The loss of interfacial coherence has a dual effect, affecting not only the aspect ratio and average growth rate of the hydrides, but also leading to redirection behaviour and an increase in the maximum stress value at the hydride tip.  As interfacial coherence loss increases, the hydride length-to-thickness ratio initially decreases ($q$ $>$ $q_c$) and subsequently increases ($q$ $<$ $q_c$), accompanied by an initial decrease and subsequent increase in the average growth rate. his phenomenon can be attributed to the anisotropic distribution of elastic interactions caused by the loss of interfacial coherence. Interestingly, the reorientation of hydrides at the mismatch degree $q_c$ remains unaffected by temperature and changes in interfacial energy.  
	
	By presenting these results, this investigation enhances our comprehension of the growth patterns and dynamic development of Zr hydrides, influenced by temperature and interfacial effects. Nevertheless, it is important to acknowledge that the above-mentioned deductions are based on theoretical assumptions of elastic energy, not accounting for plasticity. Therefore, the impact of plasticity on the outcomes calls for further inquiry.

	\section{CRediT authorship contribution statement}
	
	Zi-Hang Chen and Jie Sheng contributed equally to this work. 
	Zi-Hang Chen and Jie Sheng: Writing-Original draft preparation, Writing-Reviewing and Editing, Methodology, Investigation, Validation, Formal Analysis. 
	Yu Liu: Conceptualization, Supervision, Writing-Original draft preparation, Formal Analysis, Writing-Reviewing and Editing. 
	Xiaoming Shi: Investigation, Writing-Original draft preparation, Writing-Reviewing and Editing, Formal Analysis. 
	Houbing Huang: Conceptualization, Supervision, Writing-Reviewing and Editing.  
	Ke Xu: Investigation, Visualization, Investigation. 
	Yue-Chao Wang: Investigation, Formal Analysis.
	Shuai Wu: Investigation, Formal Analysis. 
	Bo Sun: Investigation, Formal Analysis. 
	Hai-Feng Liu: Formal Analysis, Writing-Reviewing and Editing. Hai-Feng Song: Conceptualization, Supervision, Formal Analysis, Writing-Original draft preparation and Editing.
	
	\section{Data availability}
	Data will be made availiable on request.
	
	\section{Acknowledgement}
	
	 We thank Yuan-Ji Xu, Bei-Lei Liu, Jin-Xiang Wang, Kai-Le Chen, Bo-Hao Zhao, Yu Song and Jing Wang for helpful discussions. The work was supported by the National Natural Science Foundation of China (GrantsU2230401, U1930401, 12004048, and 11971066), 
	 the National Key R\&D Program of China (Grant 2021YFB3S01S03),  the Science Challenge Project (NO. TZ2018002), and the Foundation of LCP. We thank the Tianhe platforms at the National Supercomputer Center in Tianjin.

	\section{Appendix A: Supplementary data}
	Supplementary material related to this article can be found online at...
	
	\bibliographystyle{unsrt}
	\bibliography{ref}
\end{document}


\title{Supplementary Materials to ``Phase-field simulations of the effect of temperature and interface for zirconium $\delta\mbox{-}$hydrides''}	
	
\author{Zi-Hang Chen}
\affiliation{School of Materials Science and Engineering,Beijing Institute of Technology, Beijing 100081, China}
\affiliation{Advanced Research Institute of Multidisciplinary Science, Beijing Institute of Technology, Beijing 100081, China}
\affiliation{Laboratory of Computational Physics, Institute of Applied Physics and Computational Mathematics, Beijing 100088, China}

\author{Jie Sheng}
\affiliation{Laboratory of Computational Physics, Institute of Applied Physics and Computational Mathematics, Beijing 100088, China}

\author{Yu Liu}
\email{liu\_yu@iapcm.ac.cn}
\affiliation{Laboratory of Computational Physics, Institute of Applied Physics and Computational Mathematics, Beijing 100088, China}

\author{Xiao-Ming Shi}
\affiliation{Department of Physics, University of Science and Technology Beijing, Beijing 100083, China}

\author{Houbing Huang}
\affiliation{School of Materials Science and Engineering,Beijing Institute of Technology, Beijing 100081, China}
\affiliation{Advanced Research Institute of Multidisciplinary Science, Beijing Institute of Technology, Beijing 100081, China}

\author{Ke Xu}
\affiliation{School of Materials Science and Engineering,Beijing Institute of Technology, Beijing 100081, China}
\affiliation{Advanced Research Institute of Multidisciplinary Science, Beijing Institute of Technology, Beijing 100081, China}
\affiliation{Laboratory of Computational Physics, Institute of Applied Physics and Computational Mathematics, Beijing 100088, China}

\author{Yue-Chao Wang}
\affiliation{Laboratory of Computational Physics, Institute of Applied Physics and Computational Mathematics, Beijing 100088, China}

\author{Shuai Wu}
\affiliation{School of Materials Science and Engineering,Beijing Institute of Technology, Beijing 100081, China}
\affiliation{Advanced Research Institute of Multidisciplinary Science, Beijing Institute of Technology, Beijing 100081, China}
\affiliation{Laboratory of Computational Physics, Institute of Applied Physics and Computational Mathematics, Beijing 100088, China}

\author{Bo Sun}
\affiliation{Laboratory of Computational Physics, Institute of Applied Physics and Computational Mathematics, Beijing 100088, China}

\author{Hai-Feng Liu}
\affiliation{Laboratory of Computational Physics, Institute of Applied Physics and Computational Mathematics, Beijing 100088, China}

\author{Hai-Feng Song}
\affiliation{Laboratory of Computational Physics, Institute of Applied Physics and Computational Mathematics, Beijing 100088, China}

\date{\today}	
\maketitle
	
To ensure the smoothness and continuity of the free energy function between concentration 0 and 1 at different temperatures, we employed a second-order fitting approximation of the calculation curve, guided by a thermodynamic model. The detailed procedure and methodology for this fitting process are elaborated in Sec. \ref{sec1} of our study. By utilizing this fitting method, we can accurately represent the free energy behavior and achieve a seamless transition between concentration extremes.

To calculate the strain energy associated with interface coherence during hydride formation in our modeling framework, we utilize the transformation strain derived from the Karchaturyan microelasticity theory\cite{morris2010khachaturyan}. In our stress-free transformation strain (SFTS) model, the key variable is the mismatch degree, which characterizes changes in the interface coherent state between the hydride and the matrix phase. In Section \ref{sec2}, we outline the process of calculating the characteristic strain based on directional changes, facilitating a precise characterization of the strain energy associated with hydride formation.	

\section{Quadratic polynomial fitting at different temperatures}\label{sec1}

The molar Gibbs free energy, denoted as $G_{m}^{\alpha}$ and $G_{m}^{\delta}$, is determined using the sublattice model with the Calphad method, as obtained from reference \cite{dupin1999thermodynamic}. The expressions are as follows:

The free energy density $G_{m}^{\alpha}$ and $G_{m}^{\delta}$ could be constructed based on the parabolic approximation similar to Bair \textit{et al.}'s work\cite{bair2017formation}. It is assumed that the free energy density $G_{m}^{\alpha}$ and $G_{m}^{\delta}$ could be written as: 

\begin{align}
   G_{m}^{\alpha}=a_1\left (T  \right )c_{\alpha}^{2}+b_1\left (T  \right )c_{\alpha}+c_1\left (T  \right ),\\
   G_{m}^{\delta}=a_2\left (T  \right )c_{\delta}^{2}+b_2\left (T  \right )c_{\delta}+c_2\left (T  \right ).
\end{align}

\begin{figure*}[ht]
	\centering
	\includegraphics[width=0.8\textwidth]{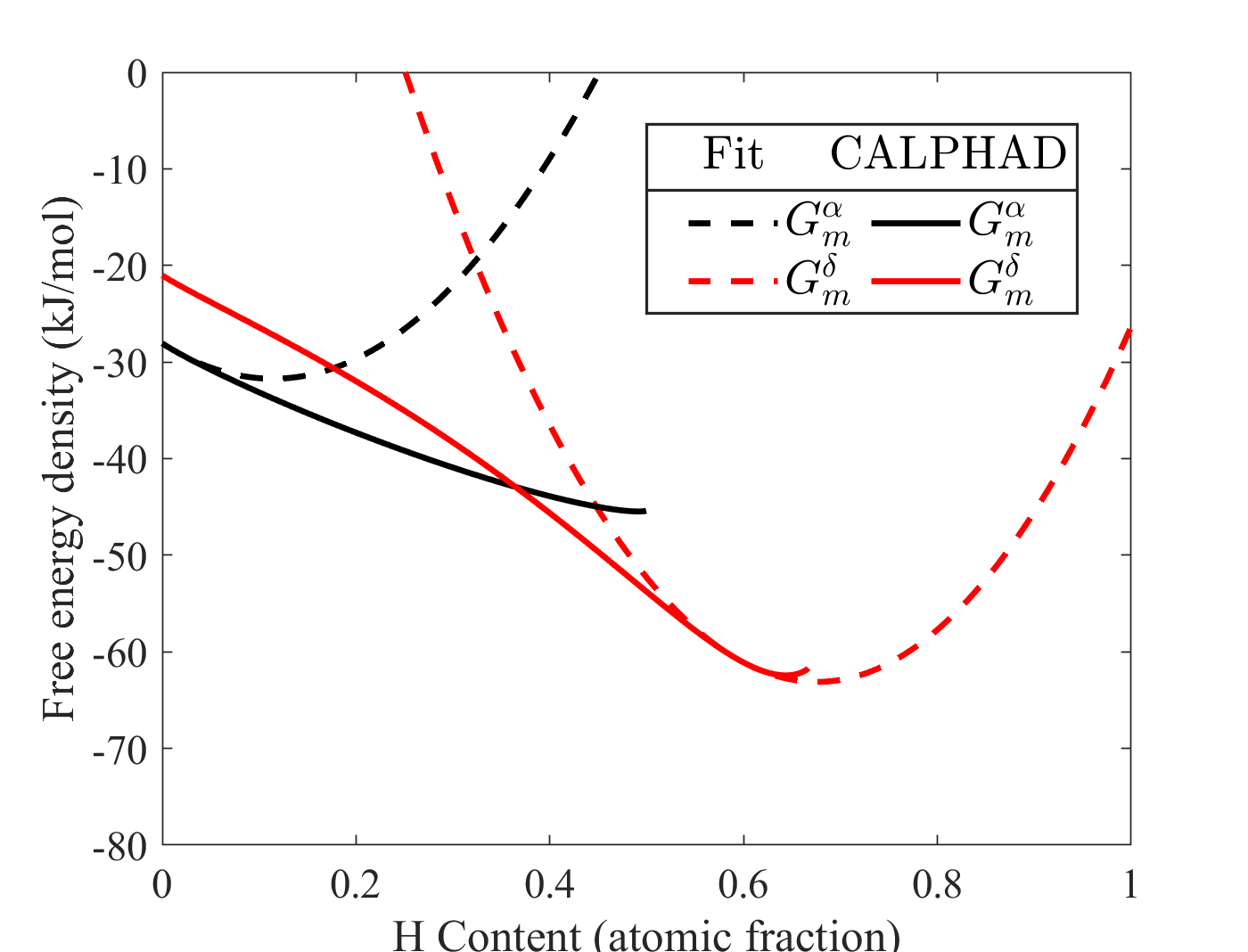}
	\caption{\label{figS1}   Parabolic approximations for molar Gibbs free energies  $G_{m}^{\alpha}$ and $G_{m}^{\delta}$ at T = 623 K based on CALPHAD.}
\end{figure*}

As an example, when calculating the two-phase free energy at a temperature of 623 K, we determine the equilibrium concentration of the Calphad free energy curve. This is accomplished by employing a second-order Taylor expansion for the Calphad-based two-phase molar Gibbs free energy expressions, denoted as $G_{m}^{\alpha}$ and $G_{m}^{\delta}$. The procedure ensures that the value of the two-phase free energy parabola expression at the equilibrium concentration point aligns with the first and second derivatives, calculated using the following formula. The resulting parabolic approximation of the Gibbs free energy in moles is illustrated in Fig. \ref{figS1}.
\begin{gather}
	f_{\alpha} \mid _{ c_{\alpha}=c_{\alpha}^{eq}}=G_{m}^{\alpha} \mid _{ c_{\alpha}=c_{\alpha}^{eq}},\\
	f_{\delta} \mid _{ c_{\delta}=c_{\delta}^{eq}}=G_{m}^{\delta} \mid _{ c_{\delta}=c_{\delta}^{eq}},\\
		{f_{\alpha}^{'}}\mid_{c_{\alpha}=c_{\alpha}^{eq}}=\frac{\partial G_{m}^{\delta} }{\partial c_{\delta} }\mid_{c_{\delta}=c_{\delta}^{eq}},\\
		{f_{\delta}^{'}}\mid_{c_{\delta}=c_{\delta}^{eq}}=\frac{\partial G_{m}^{\delta} }{\partial c_{\delta} }\mid_{c_{\delta}=c_{\delta}^{eq}},\\
		{f_{\alpha}^{''}}\mid_{c_{\alpha}=c_{\alpha}^{eq}}=\frac{\partial^2 G_{m}^{\alpha} }{\partial c_{\alpha}^{2} }\mid_{c_{\alpha}=c_{\alpha}^{eq}},\\
		{f_{\delta}^{''}}\mid_{c_{\delta}=c_{\delta}^{eq}}=\frac{\partial^2 G_{m}^{\delta} }{\partial c_{\delta}^{2} }\mid_{c_{\delta}=c_{\delta}^{eq}}. 
	\end{gather}

\section{Calculation of eigenstrain related to mismatch degree}\label{sec2}

 The lattice parameters for $\alpha$ and $\delta$ phases are employed for computing the SFTS as the following\cite{singh2007temperature}:
  
 \begin{align}
 	&a_{\alpha}=0.323118+1.6626\times 10^{-6}\times (T-298),\\
 	&c_{\alpha}=0.514634+4.7413\times 10^{-6}\times (T-298),\\
 	&a_{\delta}=0.4706+4.382\times 10^{-3}u_{\delta}^{eq}(T)+ \{ 2.475\times 10^{-6}+6.282\times 10^{-6}u_{\delta}^{eq}(T)+5.8281\times \notag\\
 	&10^{-8}\left ( {u_{\delta}^{eq}(T)}^2  \right )   \}\times (T-298),\\
 	&u_{\delta}^{eq}(T)=1.68707+\left ( -5.08641\times 10^{-4} \right )T+ \left ( 6.99511\times 10^{-7} \right )T^2+\left ( -6.61907\times 10^{-10} \right )T^3.
 \end{align}

Among them, the lattice parameters of the two phases are temperature dependent. Here, considering the lattice parameters at 623 K, we calculate:
\begin{gather}
	a_{\alpha}=0.323658,\\
	c_{\alpha}=0.516175,\\
	a_{\delta}=0.477146.
\end{gather}

\begin{figure*}[ht]
	\centering
	\includegraphics[width=0.8\textwidth]{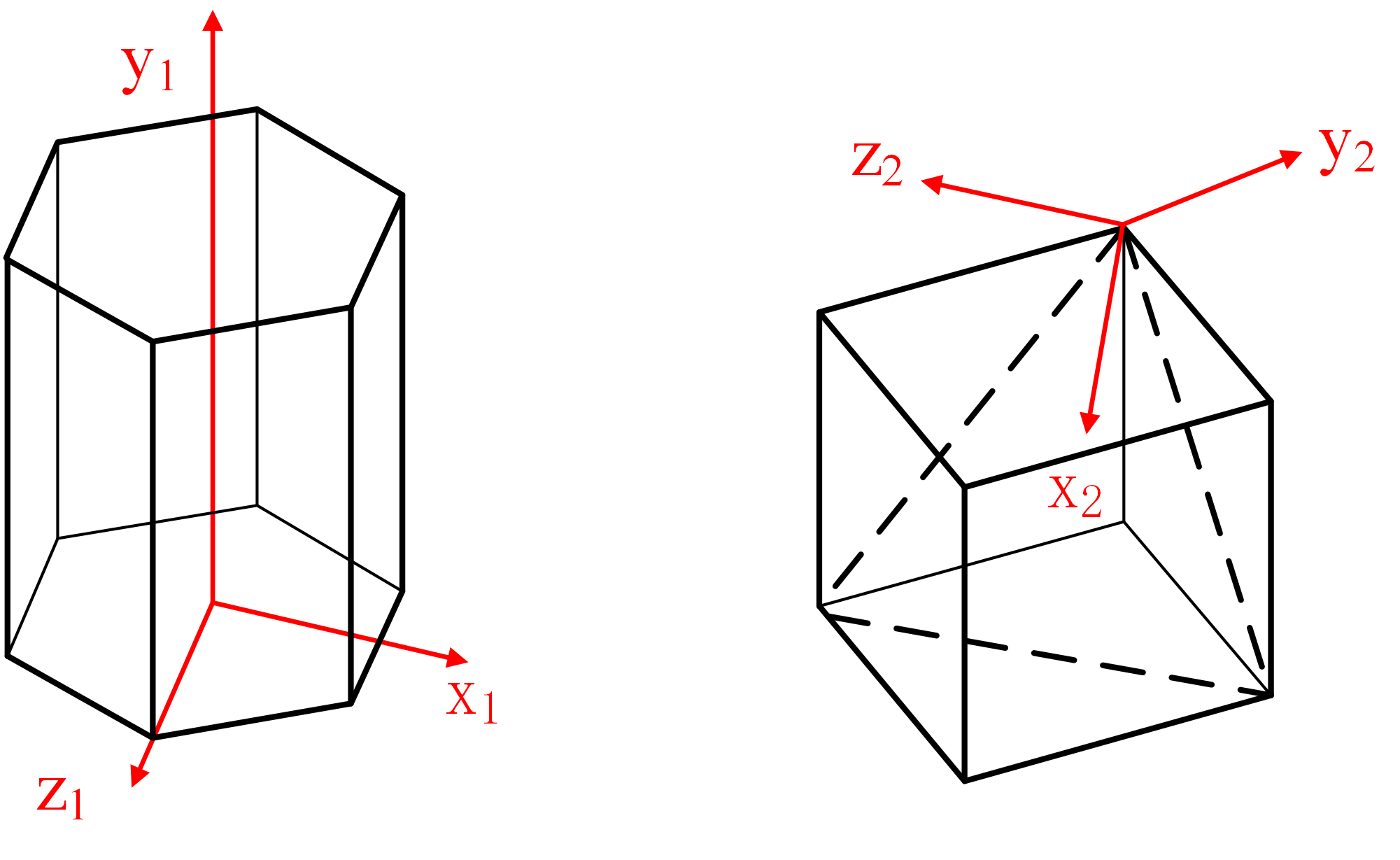}
	\caption{\label{figS2}  Coordinate systems that were used for hcp $\alpha\mbox{-}$Zr lattice and $\delta\mbox{-}\textrm{hydride}$ lattice in the current work.}
\end{figure*}

The formation of $\delta\mbox{-}\textrm{hydride}$ in $\alpha\mbox{-}$Zr involves a crystal structure transition from hcp to fcc, The orientation relation is $\left \{0001\right \}_\alpha \parallel \left \{111\right \}_\delta$ and $\left \langle 11\bar{2} 0 \right \rangle _\alpha \parallel \left \langle 1\bar{1}0 \right \rangle _\delta$. The coordinate system selected in the current work is shown in Fig. \ref{figS2}, where the coordinate system of the Zr matrix $x_1\mbox{-}y_1\mbox{-}z_1$ must be changed to the coordinate system of $\delta\mbox{-}\textrm{hydride}$ $x_2\mbox{-}y_2\mbox{-}z_2$ according to the directional relationship. For mathematical convenience and consistency, we define the deformation gradient tensor for a single strain component according to the coordinate system defined in Fig. \ref{figS2}. Deformation gradient tensor $F_1$ for shear in the $x_1$ direction, deformation gradient $F_2$ for isotropic lattice distortion in the $x_1\mbox{-}z_1$ plane, and deformation gradient tensor $F_3$ for stretching in the $y_1$ direction are expressed as follows:
\begin{align}
	F_{1}=\begin{bmatrix}
		1 & \dfrac{\dfrac{\sqrt{3}}{3}a_{\alpha } }{c_{\alpha }}&0 \\ 
		0 & 1 &0 \\ 
		0&  0& 1
	\end{bmatrix},\\
   F_{2}=\begin{bmatrix}
	\dfrac{\dfrac{\sqrt{2}}{2}a_{\delta } }{a_{\alpha }} & 0&0 \\ 
	0 & 1 &0 \\ 
	0&  0& \dfrac{\dfrac{\sqrt{2}}{2}a_{\delta } }{a_{\alpha }}
\end{bmatrix},\\
	F_{3}=\begin{bmatrix}
	1 & 0&0 \\ 
	0 & q\cdot\dfrac{\dfrac{2\sqrt{3}}{3}a_{\delta } }{c_{\alpha }} &0 \\ 
	0&  0& 1
\end{bmatrix}.
\end{align}

Among them, $a_ {\alpha} $, $c_ {\alpha} $ and $a_ {\delta} $ represents the lattice parameters of $\alpha\mbox{-}$Zr and $\delta\mbox{-}\textrm{hydride}$, respectively $ q $ is used to describe the interfacial coherence loss along the $y_1$ direction, which is called mismatch degree. We define the mismatch degree $q$ as the quantity related to $\nu$, which is expressed as: 
\begin{equation}
	\nu=\dfrac{m\cdot\dfrac{2\sqrt{3}}{3}a_{\delta }-n\cdot c_{\alpha }}{n\cdot c_{\alpha }}
\end{equation}

where $m$ and $n$ are positive integers that represent the lattice correspondence. In the limit of a perfectly coherent interface, the $\left ( m,n \right ) $ pair becomes $\left ( 1,1 \right ) $ with no interfacial dislocations. $\nu$ can be simply parameterized by the rational number $q=\left ( m,n \right ) $. Meanwhile, due to the fact that in the theory of microelasticity, regardless of the magnitude of interfacial coherence loss, the eigenstrain should ensure the same volumetric strain, an additional deformation gradient of $F$ is introduced $F_{4}$:
\begin{equation}
	F_{4}=\begin{bmatrix}
		\left (\dfrac{1}{q}  \right )^{\dfrac{1}{3}} & 0&0 \\ 
		0 & \left (\dfrac{1}{q}  \right )^{\dfrac{1}{3}} &0 \\ 
		0&  0& \left (\dfrac{1}{q}  \right )^{\dfrac{1}{3}}
	\end{bmatrix}
\end{equation}

With Lagrangian finite-strain theory\cite{lubliner2008plasticity}, the expression for the eigenstrain tensor is obtained:
\begin{equation}
	\varepsilon_{ij}^{00}=\dfrac{1}{2} \left ( F^{T}F-I \right ).
\end{equation}

In particular, while the aforementioned calculation considers shear strain, it is important to note that the path with the lowest energy for the hydride phase transition does not involve shear strain\cite{zhang2016homogeneous,louchez2017microscopic}, and this approach aligns with Simon's method\cite{simon2021investigation}. Consequently, our model does not incorporate shear strain, leading to the derivation of the following expressions for the total deformation gradient and eigenstrain:

\begin{align}
	&F=\left (\dfrac{1}{q}  \right )^{\dfrac{1}{3}}\cdot\begin{bmatrix}
		\dfrac{\dfrac{\sqrt{2}}{2}a_{\delta } }{a_{\alpha }} & 0 &0 \\ 
		0 & q\cdot\dfrac{\dfrac{2\sqrt{3}}{3}a_{\delta } }{c_{\alpha }} &0 \\ 
		0&  0& \dfrac{\dfrac{\sqrt{2}}{2}a_{\delta } }{a_{\alpha }}
	\end{bmatrix},\\
	&\varepsilon_{ij}^{00}=\frac{1}{2}\left (	\left (\dfrac{1}{q}  \right )^{\dfrac{2}{3}}\cdot
    \begin{bmatrix}\dfrac{1}{2}\cdot\left (\dfrac{a_{\delta } }{a_{\alpha }}  \right )^{2} & 0&0 \\ 
		0 & \left(\dfrac{1}{6}+\dfrac{4}{3}q^2\right)\cdot \left (\dfrac{a_{\delta } }{c_{\alpha }}  \right )^{2} &0 \\ 
		0&  0& \dfrac{1}{2}\cdot\left (\dfrac{a_{\delta } }{a_{\alpha }}  \right )^{2}
	\end{bmatrix}-I  \right ).
\end{align}

Therefore, under 623 K, the tensor form of the eigenstrain is written as a matrix related to $q$:
\begin{equation}
	\varepsilon_{ij}^{00}=\frac{1}{2}\left (\left (\dfrac{1}{q}  \right )^{\dfrac{2}{3}}\cdot
	\begin{bmatrix}
		1.08667 & 0 &0 \\ 
		0 & \left(\dfrac{1}{6}+\dfrac{4}{3}q^2\right)\cdot0.85449 &0 \\ 
		0&  0& 1.08667
	\end{bmatrix}-I\right ).
\end{equation}

Specifically, in the simulation, the effect of shear strain on the morphology is mainly in the torsion effect. In our study, the effect of shear strain is ignored in order to facilitate numerical calculation. Only the eigenstrain in the 11 and 22 directions are considered, so the calculation formula is:
\begin{gather}
	\varepsilon_{11}^{00}=\dfrac{1}{2}\left [ \left (\frac{1}{q} \right )^{\frac{2}{3}}\cdot1.08667-1\right ],\\
	\varepsilon_{22}^{00}=\dfrac{1}{2}\left [ \left (\frac{1}{q} \right )^{\frac{2}{3}}\cdot \left(\dfrac{1}{6}+\dfrac{4}{3}q^2\right)\cdot0.85449-1\right ].
\end{gather}

\bibliographystyle{unsrt}
\bibliography{ref}
\clearpage